\documentclass[a4paper, amsfonts, amssymb, amsmath, reprint, showkeys, nofootinbib,superscriptaddress, twoside,aps,prd,longbibliography]{revtex4-1}
\pdfoutput=1
\usepackage[english]{babel}
\usepackage[utf8]{inputenc}
\usepackage[colorinlistoftodos, color=green!40, prependcaption]{todonotes}
\usepackage{amsthm}
\usepackage{mathtools}
\usepackage{physics}
\usepackage{xcolor}
\usepackage{graphicx}
\usepackage[left=23mm,right=13mm,top=35mm,columnsep=15pt]{geometry} 
\usepackage{adjustbox}
\usepackage{placeins}
\usepackage[T1]{fontenc}
\usepackage{lipsum}
\usepackage{csquotes}
\usepackage{hyperref}
\hypersetup{  
    colorlinks=true,                
    breaklinks=true,                
    urlcolor= black,                
    linkcolor= black,                
    bookmarksopen=false,
    filecolor=black
    citecolor=black,
    linkbordercolor=black
}
\hypersetup{  
    colorlinks=true,                
    breaklinks=true,                
    urlcolor= black,                
    linkcolor= blue,                
    bookmarksopen=false,
    filecolor=blue,
    citecolor=green,
    linkbordercolor=blue
}

\newcommand{\beginsupplement}{%
        \setcounter{figure}{0}%
        \renewcommand{\thefigure}{S\arabic{figure}}%
        \setcounter{equation}{0}%
        \renewcommand{\theequation}{S\arabic{equation}}%
}

\usepackage{units}
\usepackage{soul}
\usepackage{color}
\usepackage{changes}
\usepackage{subcaption}
\usepackage{diagbox}
\usepackage{stix}
\usepackage{ragged2e}
\usepackage{caption}
\hypersetup{pdfauthor={A. Jaroš, M.S. Seifner, J. Toyfl, B. Czasch, I.C. Bicket, P. Haslinger}, pdftitle={Sensing Spin Systems with a Transmission Electron Microscope}}

\begin{document}
    \title{Sensing Spin Systems with a Transmission Electron Microscope}

\author{Antonín Jaroš}
    \affiliation{Vienna Center for Quantum Science and Technology, Atominstitut, USTEM, Technische Universität Wien, Vienna, Austria}
\author{Michael S. Seifner}
    \affiliation{Vienna Center for Quantum Science and Technology, Atominstitut, USTEM, Technische Universität Wien, Vienna, Austria}
\author{Johann Toyfl}
    \affiliation{Vienna Center for Quantum Science and Technology, Atominstitut, USTEM, Technische Universität Wien, Vienna, Austria}
\author{Benjamin Czasch}
    \affiliation{Vienna Center for Quantum Science and Technology, Atominstitut, USTEM, Technische Universität Wien, Vienna, Austria}
\author{Isobel C. Bicket}
    \affiliation{Vienna Center for Quantum Science and Technology, Atominstitut, USTEM, Technische Universität Wien, Vienna, Austria}
\author{Philipp Haslinger}
    \affiliation{Vienna Center for Quantum Science and Technology, Atominstitut, USTEM, Technische Universität Wien, Vienna, Austria}

\date{\today}
\begin{abstract}
We present a novel method that combines spin resonance spectroscopy with transmission electron microscopy (TEM), enabling localized \textit{in situ} detection of microwave (MW)-driven spin excitations. Our approach utilizes continuous wave MW excitation at GHz frequencies, while employing the free-space electron beam as a signal receiver to sense spin precession. Spin state polarization is achieved via the magnetic field of the TEM's polepiece, while a custom-designed microresonator integrated into a TEM sample holder drives spin transitions and modulates the electron beam. This modulation enables phase-locked detection with picosecond temporal resolution, allowing the isolation of spin precession contributions to the electron beam deflection with a sensitivity of $\sim 280$ prad. The presented technique lays foundations for the MW spectroscopic \textit{in situ} exploration of spin dynamics at the nanoscale.
\end{abstract}

\maketitle

The microwave (MW) frequency band plays a crucial role in many scientific fields. In atomic and molecular quantum optics, it enables the coherent manipulation of long-lived transitions, such as the cesium hyperfine transition which defines the second \cite{bochmann2010lossless, wynands2005atomic, viteau2008molecules, porterfield2019MW}. MW spectroscopy has also proved disruptive in other disciplines such as condensed matter physics, chemistry, biology and medicine, where electron spin resonance (ESR) \cite{bienfait2016reaching, jarovs2024electron}, nuclear magnetic resonance \cite{Callaghan1993Principles, Boero2003} and ferromagnetic resonance \cite{wang2018fmr, serha2024magnetic} are key to advancements in non-invasive quantum technologies, including magnetic resonance imaging~(MRI)~\cite{vlaardingerbroek2013MRIbook}.

Such technologies rely on sensing a fundamental quantum property of matter: the spin. MW radiation is used to coherently manipulate spin states, which exhibit surprisingly long coherence times~\cite{Callaghan1993Principles}. 
The chemical environment can alter these spin states, providing spectroscopic insights into the atomic structure of a given sample. However, conventional spectroscopic tools typically average over macroscopic samples. Enhancing the sensitivity and spatial resolution of these techniques is crucial for advancing the study of spin ensembles down to the atomic scale. So far, efforts to sense spins on the nanoscale are limited to specific sample geometries \cite{simpson2017electron, seifert2020single}, or equipment only hosted at large-scale facilities \cite{sluka2019emission, wintz2016magnetic, dieterle2019coherent}. 
However, since the spin is linked to the Bohr magneton, any changes to the spin state alter the local magnetic field, which can be detected by electrons~\cite{Haslinger2024}, offering a potential pathway for more accessible spin detection techniques.

Transmission electron microscopy (TEM) is a powerful technique that enables atomic-scale investigations using a highly controlled free-space electron beam \cite{Reimer2008Transmission, sawada2009stem47pm, Erni2009Atomic-resolution}. Ultra-fast TEMs with laser triggered sources \cite{Houdellier2018, Feist2017Ultrafast, morimoto2018diffraction, Barwick2009Photon, wang2020coherent, alcorn2023EMchemicaldynamics} or chopped electron beams \cite{verhoeven2018high, van2018dual} extend those capabilities by probing dynamic processes on femto- and even atto-second timescales. For detailed spectroscopic specimen analysis, TEM offers an advanced suite of analytical techniques such as electron energy loss spectroscopy~\cite{egerton2011electron}.
Recent advancements allow for the detection of energy losses and gains associated with optical interactions~\cite{Polman2019Electron-beam, Barwick2009Photon, auad2023muev}, atomic vibrations (phonons) and mid-IR plasmonic excitations \cite{lagos2022HREELS_review, kumar2022vibEM}. Nevertheless, the MW frequency band is not accessible in EELS due to the energy width of the primary electron beam. Consequently, new techniques need to be developed to study phenomena in this frequency regime.

Recent research efforts focus on the development of custom-made sample holders capable of exciting samples at MW frequencies \cite{harvey_ultrafast_2021, pollard_direct_2012, moller_few-nm_2020, jarovs2024electron, Goncalves2017}. Stroboscopic MW-pump electron-probe schemes were utilized, enabling, e.g., ultra-fast imaging of magnetic vortex cores \cite{wesels2022beamdeflections, pollard_direct_2012, moller_few-nm_2020}, probing of beam deflections near MW circuits \cite{wesels2022beamdeflections, Goncalves2017}, and more recently, the visualization of spin waves in ferromagnetic thin films \cite{liu2025correlated}. In prior work, we integrated a specially designed MW resonator on a custom TEM sample holder, allowing for conventional ESR investigation of miniaturized sample sizes \cite{jarovs2024electron}. This setup realizes coherent spin manipulation at frequencies around 4.7 GHz, corresponding to an excitation energy of $\approx 20$ µeV.

Using the same setup, we excite spin states with MWs and read out the response using the free-space electron probe, which acts as a localized receiver. 
Since the electron beam is also modulated by the continuous-wave (CW) MW excitation field, we can achieve phase-locked detection with a temporal resolution in the picosecond range. This method allows us to isolate and study precession-induced shifts visible in the electron beam's angular distribution in momentum space, with picoradian (prad) sensitivity.
The developed MW-pump electron-probe scheme allows for the direct investigation of spin signatures via the electron beam, which we refer to as SPINEM (SPIN Electron Microscopy), drawing a parallel to the established technique of photon-induced near-field electron microscopy (PINEM) \cite{Barwick2009Photon,lummen2016imaging}. 
Further improvements of the presented technique could enable investigations of coherently driven spin phenomena at the atomic scale.

\begin{figure*}
	\centering
	\includegraphics[width=1\textwidth]{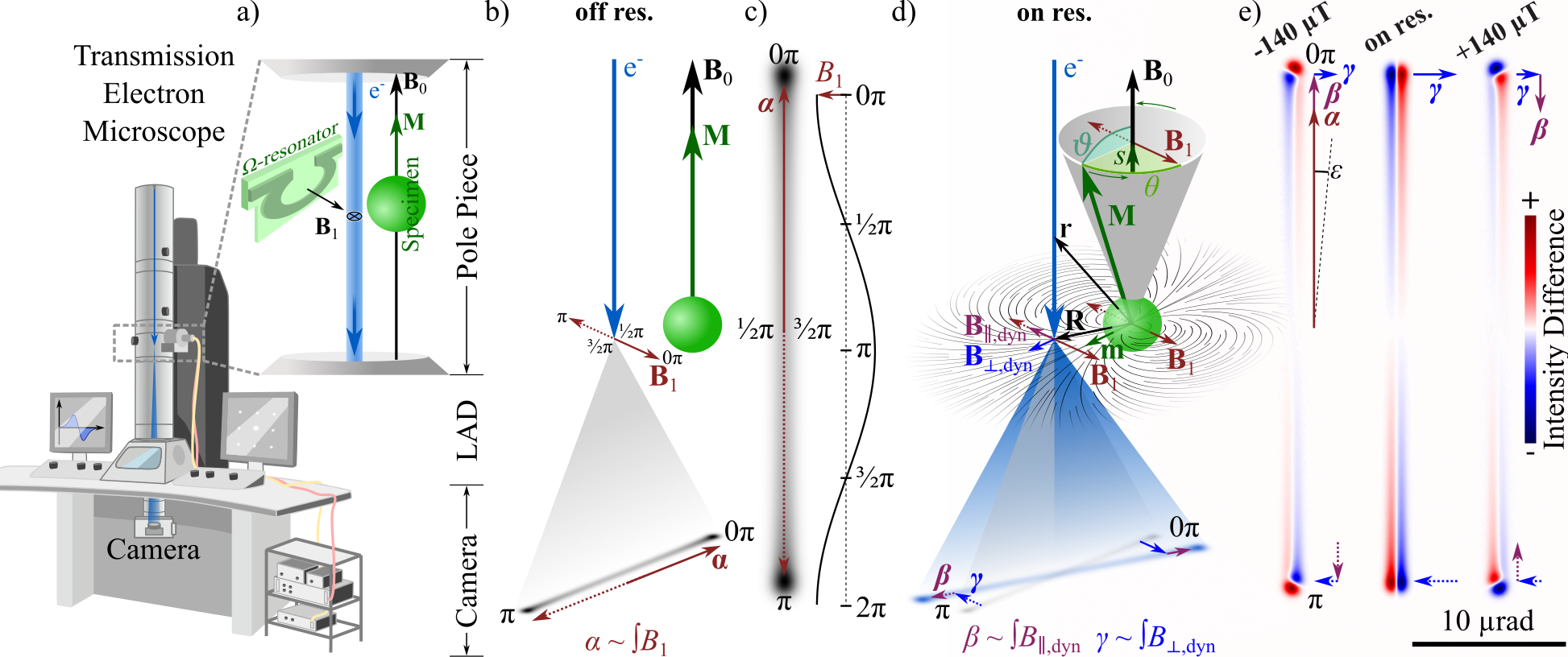}
	\caption{
    \justifying
    Continuous SPINEM \textit{in situ} experiment in a TEM. (a) A specimen with addressable spin states is placed in the static magnetic field $\mathbf{B}_0$ of the objective lens pole piece. The sample's magnetization $\mathbf{M}$ aligns with $\mathbf{B}_0$, resulting in minimal electron probe interaction and no angular beam deflection. (b) When MWs are introduced but are far off-resonant, their dynamic $\mathbf{B}_1$ field deflects the electron beam due to the Lorentz force by $\boldsymbol{\alpha}$ without perturbing the spin state alignment with $\textbf{B}_0$. (c) Calculation of the beam profile, which is a time-averaged projection of the sinusoidal beam deflection $\alpha(\omega,t) \sim \int B_1(z,\omega, t) \sim \cos(\omega t)$, where $\omega t$ is the phase of the driving microwaves. (d) At resonance, the spin system is dynamically driven by the $\mathbf{B}_1$ field and the magnetization $\mathbf{M}$ begins to precess at an angle $\vartheta$ around $\mathbf{B}_0$, while lagging by a phase $\theta$ behind the driving field. This results in an out-of-plane static component $\mathbf{s}(\omega)$ and an in-plane dynamic component $\mathbf{m}(\omega,t)$. Additional beam deflection both perpendicular ($\beta(\mathbf{R},\omega,t) \sim \int B_{\parallel,\text{dyn}} (\textbf{r}, \omega, t)$) and parallel ($\gamma(\mathbf{R},\omega,t) \sim \int B_{\perp,\text{dyn}} (\textbf{r}, \omega, t) $) to $\alpha$ can be observed. These deflections are induced by the dynamic fields $B_{\parallel,\text{dyn}}$ and $B_{\perp,\text{dyn}}$, generated by the in-plane magnetization $\mathbf{m}$ at the electron probe position $\mathbf{r}=(x,y,z)$. (e) Difference images referenced to the beam deflection in (c) revealing that $\mathbf{m}$ causes the beam profile to extend, compres, and tilt in the camera plane as a function of the detuning from resonance. The calculations are performed for a spherical specimen with a $75$ \textmu m radius and a spin density of $1.5 \text{ spins/nm}^3$ (thermally polarized), using a point-like electron probe positioned $150$ \textmu m from the sphere center. The magnetic fields are $B_0 = 0.17$ T and $B_{1,max} = 20$ \textmu T.}
    \label{fig:1} 
\end{figure*} 

\section*{Experimental Setup: SPINEM}

A conventional ESR setup, similar to those in \cite{Boero2003, narkowicz2005planar, narkowicz2008scaling}, along with a custom-built sample holder \cite{jarovs2024electron} is used to perform the SPINEM experiments. The holder features an $\Omega$-shaped microresonator, impedance matched to $\sim$4.7 GHz. The electron spin-active specimen ($\alpha, \gamma$-bisdiphenylene-$\beta$-phenylallyl, BDPA, diameter $\sim100$ \textmu m) is positioned near the center of the microresonator, while ensuring sufficient space to perform aloof-mode measurements, as illustrated in Fig.~\ref{fig:1}a. We operate the TEM in low magnification mode, producing a $B_0$ field in the range of 170~mT at the specimen, which is used for spin polarization and tuning the resonance condition. The magnetization vector of the spin states $\mathbf{M}$ aligns with the $\mathbf{B}_0$ field, resulting in only minimal electron beam deflections.

Irradiating the sample with a time-dependent magnetic field $\mathbf{B}_1(t)$ orthogonal to the static spin polarization induces dynamic spin evolution. The coherent interaction between $\mathbf{B}_1(t)$ and the sample's spin states causes a precession of the spins and corresponding magnetization. The magnetization follows the Bloch equations \cite{Callaghan1993Principles}, which capture the characteristic dynamics of spin resonance spectroscopy. Solving these equations yields the resonance condition $\omega_{\text{res}} = 2\pi \Gamma B_0$, where $\Gamma$ is the gyromagnetic ratio.

When a CW off-resonant $\mathbf{B}_1(z,\omega,t)=\mathbf{B}_{1,\text{max}}(z,\omega) \cos(\omega t)$ field is applied, $\mathbf{M}$ remains parallel to $\mathbf{B}_0$, while the beam experiences a total angular deflection $\boldsymbol{\alpha} (\omega, t)$ due to the dynamic Lorentz force, integrated across the electron's trajectory, as illustrated in Fig.~\ref{fig:1}b:
\begin{equation}
\boldsymbol{\alpha} (\omega, t) \sim 
\int_{-\infty}^{+\infty}
\mathbf{v}_{\text{e}} \times \mathbf{B}_{1,\text{max}}(z, \omega) \cos(\omega t) \,dz,
\end{equation}
\noindent
where $\mathbf{v}_{\text{e}}$ is the velocity vector of the electron beam and the $z$-axis aligns with the optical axis. As the microresonator spans a large portion of the accessible sample region, the $B_1(z,\omega,t)$ amplitude exhibits only minor variations in the $xy$-specimen plane, which we neglect here.

The deflection amplitude $\alpha_{\text{max}} (\omega)$ depends on the $B_{1}(\omega)$ driving field, therefore reflecting the microresonator impedance match and MW power delivery to the location of the electron probe, see Fig. \ref{fig:vna_tem} in supplementary information (SI) \hyperref[sec:AppendixA]{a}.
Fig.~\ref{fig:1}c shows a calculated pattern with $\alpha_{\text{max}} = 16.75$ \textmu rad as a result of a homogeneous driving field $B_{1,\text{max}} = 20$~\textmu T within the microresonator. 

The recorded electron deflection pattern represents a time-averaged projection of the sinusoidally modulated electron beam.
Without ultra-fast TEM capabilities, we are not able to temporally resolve the $1/4.7~\text{GHz}^{-1}~\approx~210$ ps period of the oscillating electron probe, necessitating the development of analysis strategies for the time-averaged pattern.
However, different points along the pattern correspond to different $\omega t$ phases of the driving field, effectively creating a position-encoded time resolution, similar to Lissajous curves \cite{borrelli2024direct}.

\begin{figure}
    \centering
    \includegraphics[width=0.48\textwidth]{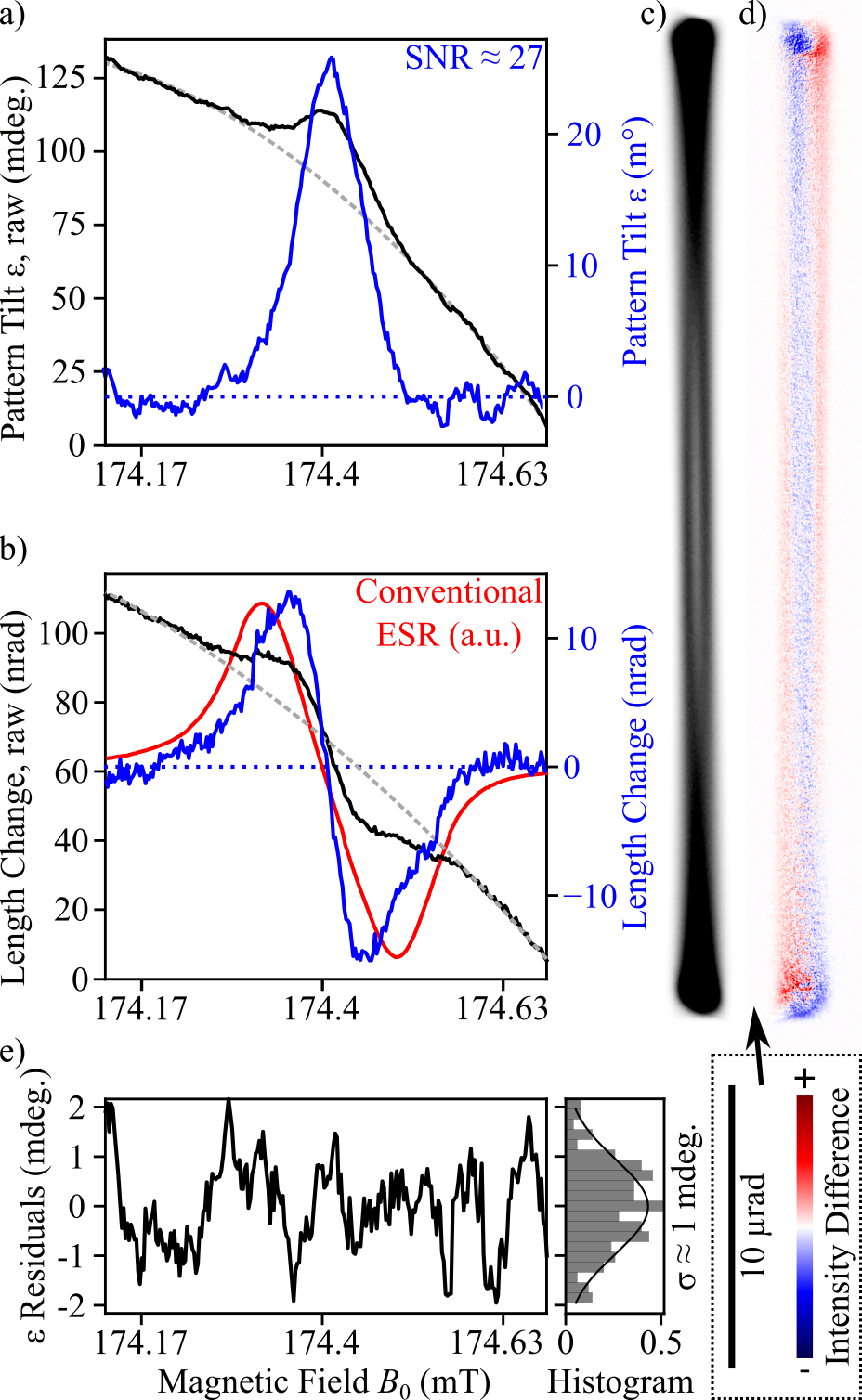}
    \caption{
    \justifying 
    CW SPINEM measurement. (a) Pattern tilt $\varepsilon$ represents ESR absorption spectra measurements. (b) Pattern length change corresponds to the detection of ESR dispersion spectra. For (a) and (b), measured data is shown in black, the fitted quadratic polynomial offset is displayed in dashed gray, and the final signal in blue. For reference, a conventional ESR absorption spectrum measurement is shown in (b) (red). 
    The difference image (d), referenced to the far off-resonance image (c), highlights the tilting $\varepsilon$ at resonance, with positive and negative intensity differences shown in red and blue, respectively. (e) Fit residuals of pattern tilt $\varepsilon$, following a Gaussian distribution with a standard deviation of 1 millidegree (mdeg.). This corresponds to an SNR of 27. Experimental conditions: MW frequency $\nu = 4.89$ GHz, with a source output power $P_{g} = 24$ dBm. Conventional ESR measurement: lock-in time constant $\tau = 30$ ms, modulation frequency 101.01 kHz, and modulation field 165 µT. SPINEM measurement: camera length 600 m, camera acquisition time 20 s per spectral point, beam current 500 pA.}
    \label{fig:2}
\end{figure}

\begin{figure*}
    \centering
    \includegraphics[width=1\textwidth]{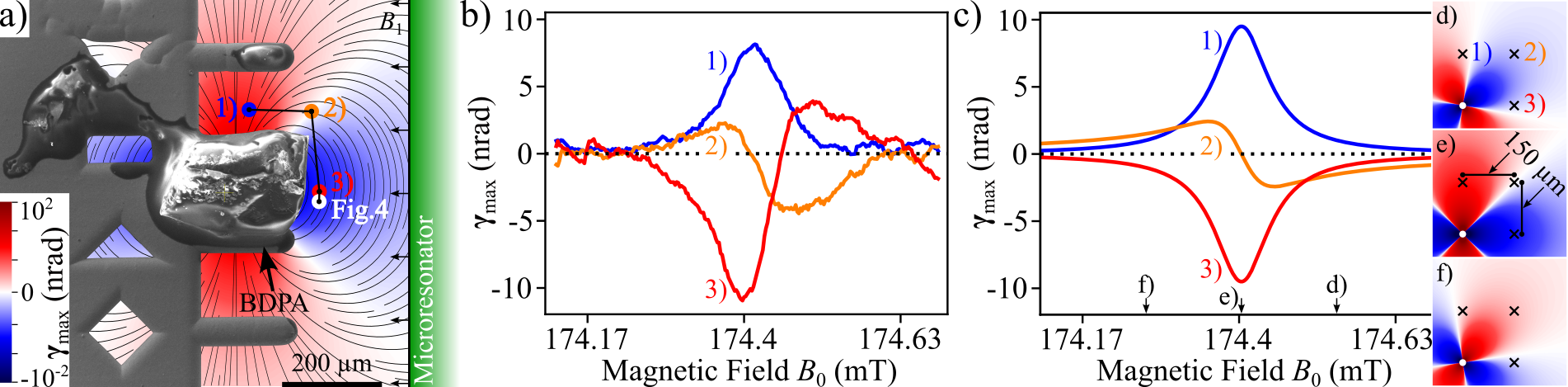}
    \caption{
    \justifying 
Spatially resolved SPINEM. (a) SEM image of specimen mounted on a FIB lift-out grid. The microresonator position is highlighted in green. The image is overlaid with exemplary magnetic dipole field lines of the specimen and with calculated beam deflection $\gamma_{\text{max}}$ at resonance and for MW phase $\omega_{\text{res}}t = 0$.  (b) SPINEM spectra for different probe positions (1-3). The detected tilt angle is converted into beam deflection $\gamma_{\text{max}}$, directly linked to the spin-electron interaction. Positions (1) and (3) exhibit lineshapes similar to conventional ESR absorption spectra, whereas position (2) is reminiscent of ESR dispersion spectra. Since all measurements are referenced to the same $B_1$ field from the driving microwaves, the beam deflection $\gamma_{\text{max}}$ serves as a measure of the in-plane magnetization of the specimen, perpendicular to the polarization of the driving microwaves, as the $B_0$ biasing field varies. (c) Calculations for $\gamma_{\text{max}}$, for positions (1-3), chosen to reflect the experimental arrangement in (b). Overall, the experiment and calculations are in good agreement. Panels (d-f) show 2D maps of calculated $\gamma_{\text{max}}$ at -140 \textmu T, 0 \textmu T, and +140 \textmu T $B_0$ detuning from resonance. MW generator output power was set to $P_{g} = 24$ dBm (1), $P_{g} = 22$ dBm (2) and $P_{g} = 21$ dBm (3). Other parameters as in Fig. \ref{fig:2}}
    \label{fig:3}
\end{figure*}

For CW MW excitation at and near the resonance, $B_1(\omega\approx\omega_{\text{res}})$, the deflection pattern is imprinted with the phase-locked spin signature.
Additional beam deflections both parallel $\boldsymbol{\beta} (\mathbf{R}, \omega, t)$ and perpendicular $\boldsymbol{\gamma} (\mathbf{R}, \omega, t)$ to $\boldsymbol{\alpha}(\omega,t)$, arise due to the dynamic in-plane component $\mathbf{m}(\omega, t)$ of the precessing magnetization $\boldsymbol{M}(\omega, t)$. Here, $\mathbf{R}=(x,y)$ represents the position of the electron probe relative to the specimen, in the specimen plane. A theoretical derivation of $\boldsymbol{\beta}$ and $\boldsymbol{\gamma}$ deflections is presented in SI \hyperref[sec:AppendixB]{b}. The deflections can be expressed by:
\begin{equation}\label{beta}
\boldsymbol{\beta} (\textbf{R},\omega ,t) \sim 
\int_{-\infty}^{+\infty}
\mathbf{v}_{\text{e}} \times \mathbf{B}_{\parallel,\text{dyn}}(\textbf{r},\omega, t) \,dz,
\end{equation}
\begin{equation}\label{gamma}
\boldsymbol{\gamma} (\textbf{R},\omega ,t) \sim 
\int_{-\infty}^{+\infty}
\mathbf{v}_{\text{e}} \times \mathbf{B}_{\perp,\text{dyn}}(\textbf{r},\omega, t) \,dz,
\end{equation}
\noindent where $\mathbf{B}_{\parallel,\text{dyn}}(\textbf{r},\omega, t)$ and $\mathbf{B}_{\perp,\text{dyn}}(\textbf{r},\omega, t)$ denote, respectively, the collinear and perpendicular components of the specimen's magnetic field relative to the driving field $\mathbf{B}_1(z,\omega,t)$ at the electron probe position $\mathbf{r}=(x,y,z)$, as illustrated in Fig.~\ref{fig:1}d. The fields vary with detuning from resonance as the magnetization $\mathbf{M}(\omega,t)$ precesses at an angle $\vartheta$ relative to the static (z-component) spin alignment $\mathbf{s}(\omega)$, lagging by a phase $\theta$ behind the driving field phase $\omega t$.
Considering a swift interaction between the fast electron and the sample of effectively less than 5 ps, the magnetization precession dynamics ($210$ ps period) can be considered effectively static. Across resonance, $\boldsymbol{\beta}$ and $\boldsymbol{\gamma}$ are expected to reach a maximum amplitude of $\sim10$ nrad, calculated for $B_{1,\text{max}} = 20$~\textmu T and a spin polarization of about $4\cdot10^{-4}$ at 290~K and $B_0=170$~mT.

The phase-encoded deflections $\boldsymbol{\beta}(\mathbf{R},\omega,t)$ and $\boldsymbol{\gamma}(\mathbf{R},\omega,t)$ lead to an electron redistribution. Calculated patterns at a specific electron probe position $\mathbf{R}$, visualized in Fig.~\ref{fig:1}e, illustrate the effect of $\boldsymbol{\beta}$, showing an expansion of the electron pattern at a $B_0$ detuning of -140 \textmu T and a compression at +140 \textmu T from resonance. Simultaneously, the pattern tilts by an angle $\varepsilon$ around the optical axis as a result of $\boldsymbol{\gamma}$, with the largest pattern tilt occurring at resonance. 

The images in Fig.\ref{fig:1}e are difference images, obtained by subtracting the electron pattern at resonance from the far off-resonance pattern shown in Fig.~\ref{fig:1}c. Measuring the deflections $\boldsymbol{\beta}(\mathbf{R},\omega,t)$ and $\boldsymbol{\gamma}(\mathbf{R},\omega,t)$ provides information about the dynamic field generated by the in-plane magnetization $\mathbf{m}(\omega,t)$ (see eq.~\ref{beta},~\ref{gamma}). Assuming a point-like specimen and electron probe, which is positioned at $\textbf{R}$ near the specimen, it can be shown that $\boldsymbol{\beta}$ and $\boldsymbol{\gamma}$ correspond to the absorption and dispersion spectra of a conventional ESR, see~SI~\hyperref[sec:AppendixB]{b}.

Figure \ref{fig:2} illustrates our SPINEM measurement technique, employing a CW MW-pump and electron-probe scheme. The electron probe is placed near the edge of the specimen in aloof mode. 
The objective lens (OL) excitation is varied to tune the $B_0$ field while keeping the MW excitation frequency constant to perform a magnetic field scan over the resonance. 

Following the data processing steps detailed in the Methods, Fig.~\ref{fig:2}a presents the extracted SPINEM pattern tilt $\varepsilon$ (black curve). Varying the OL excitation induces a background tilt, which is accounted for by fitting and subtracting an offset (dotted curve). The resulting signal (blue curve) is similar to conventional (lock-in) ESR absorption spectra. At resonance, we extract a maximum pattern tilt of $\varepsilon \approx 26$ millidegree, caused by the spin precession. The effective change of the pattern length, correlated with the $\beta$ parameter, is shown in Fig.~\ref{fig:2}b, revealing strong similarities to conventional ESR dispersion spectra. We observe a maximum pattern elongation of approximately 15 nrad. For reference, we measured an \textit{in situ} conventional lock-in ESR spectrum (in red) of the same specimen immediately following the SPINEM measurement. Note that lock-in detection captures the derivative of the absorption signal, producing a profile visually similar to ESR dispersion spectra. Our SPINEM measurement matches well with the conventional ESR reference in both resonance frequency and line width. 
The conventional measurement exhibits a full width at half maximum (FWHM) of $5.6\pm0.1$ MHz due to broadening caused by the field modulation amplitude $B_m$ of 165 µT. 
In contrast, the SPINEM absorption spectrum reaches a FWHM of $3.1\pm0.2$ MHz.

Electron beam patterns, such as the far off-resonance case in Fig. \ref{fig:2}c, are captured at a camera length of 600 m in low-angle diffraction (LAD) mode, reaching a maximum deflection $\alpha_{\text{max}}$ of 16.65 µrad. The difference image in Fig. \ref{fig:2}d, referenced to Fig. \ref{fig:2}c, visualizes the pattern tilt $\varepsilon$ at resonance. Positive and negative intensity differences are indicated in red and blue, respectively. A tilt of 26 millidegrees results in a beam displacement of 0.7 pixels at the pattern endpoint, or equivalently, $\gamma_{\text{max}} = \alpha_{\text{max}} \tan{(\varepsilon)} = 7.5$ nrad. This deflection is a direct consequence of the spin-electron interaction.

Fig. \ref{fig:2}e presents the residuals from the offset fit of pattern tilt $\varepsilon$, which reveals a standard deviation of 1 millidegree, corresponding to a $\gamma$ deflection uncertainty of 280 prad and a beam displacement sensitivity of $\sim 1/40$ of a pixel on the Gatan Rio 4K camera. This represents a signal-to-noise ratio (SNR) of $\approx27$ in our measurement.

The beam spot size in the sample plane is approximately 30 \textmu m, limiting the spatial resolution of our SPINEM experiments. Figure \ref{fig:3}a presents a scanning electron microscope (SEM) image, marking the probed positions (1–3) examined using the SPINEM technique. By evaluating the beam deflection $\boldsymbol{\gamma}(\mathbf{R},\omega,t)$, we sense the dynamic $B_{\perp,\text{dyn}}(\mathbf{r},\omega,t)$ field generated by the in-plane magnetization $\mathbf{m}(\omega,t)$ at each probe position $\mathbf{R}$. Position (1) corresponds to the same measurement as shown in Fig. \ref{fig:2}. 

The extracted spectra for the other positions qualitatively reproduce the conventional ESR spectra, showing the expected dispersion (2) and absorption (3) features, see Fig. \ref{fig:3}b. At resonance ($B_0= 174.4~\text{mT}$), it is apparent that the beam deflection $\gamma_{\mathrm{max}}$, which is proportional to $B_{\perp,\text{dyn}}$, is maximal for (1) and (3), but is zero for (2). Note that $\boldsymbol{\gamma}^{(3)} = - \boldsymbol{\gamma}^{(1)}$, therefore $\mathbf{B}_{\perp,\text{dyn}}^{(3)} = - \mathbf{B}_{\perp,\text{dyn}}^{(1)}$. As expected, far off resonance, the in-plane dynamic field $B_{\text{dyn}}$ decays to zero for all positions. For visualization, Fig. \ref{fig:3}a is overlaid with the calculated deflection $\gamma$ for a MW phase of $\omega_{\text{res}} t = 0, \pm 2\pi,\dots$. For our calculations, we assume a point-like electron probe and spin ensemble, see SI~\hyperref[sec:AppendixB]{b} for more details. 

Fig \ref{fig:3}c shows calculations of $\gamma$ across the resonance at positions (1-3), chosen to reflect the effects on the electron beam as shown in Fig. \ref{fig:3}b. 
Although detailed information on environmental effects and the sample’s microscopic composition, size, and radiation-induced damage is missing, the calculations show excellent agreement with the experimental data. Intensity variations in Fig. \ref{fig:3}b can be attributed to changes in the input MW power (e.g., $B_1$). At position (1), where the electron probe is furthest from the microresonator, the deflection $\alpha_{\text{max}}$ is weakest, which is compensated by increasing the MW power to optimize sensitivity in evaluating the tilt $\varepsilon$. 

Analogously to the calculated 2D map presented in Fig. \ref{fig:3}a, further calculated maps are plotted at $B_0$ detunings of -140~\textmu T, 0~\textmu T, and +140 \textmu T from resonance, see Fig. \ref{fig:3}d-f. These maps display variations in both magnetization amplitude $m$ and phase $\theta$. The latter reflects the degree to which the spin system magnetization $\mathbf{m}$ lags behind the driving MW field $\mathbf{B}_1$.

\begin{figure}
    \centering
    \includegraphics[width=0.48\textwidth]{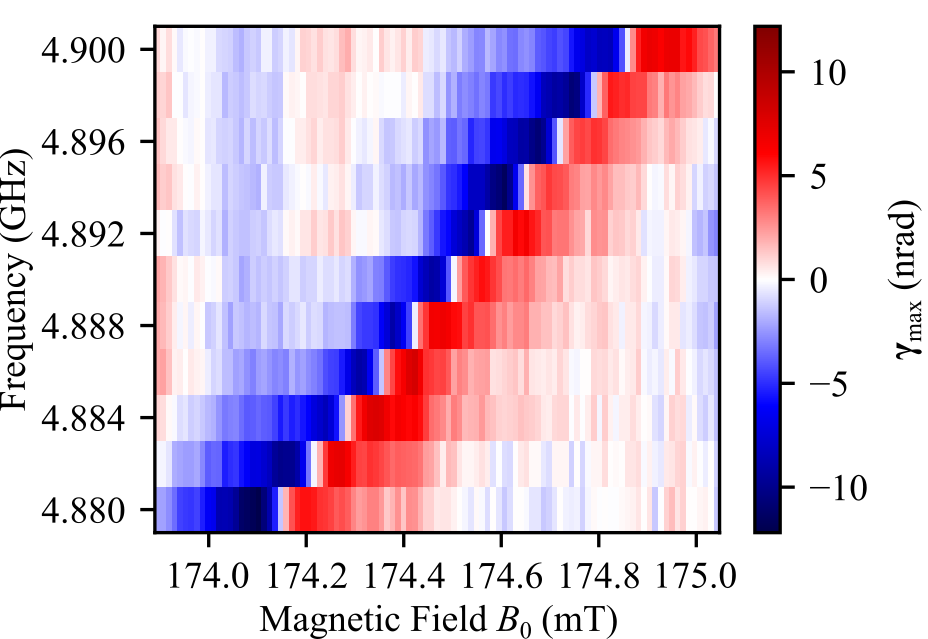}
    \caption{
    \justifying 
Consecutive field and frequency MW-pump, electron-probe measurement. The signal exhibits a linear trend, consistent with the expected Zeeman splitting of the spin states: $\omega = 2\pi \Gamma B_0$, where $\Gamma = (27.9 \pm 1.2)$. Experimental settings: MW output power $P_{g} = 21$ dBm.}
    \label{fig:4}
\end{figure}

Fig. \ref{fig:4} represents the electron beam deflection $\gamma_{\text{max}}$ in consecutive frequency and field sweeps. This measurement illustrates our ability to precisely control and detect the spin state. As the biasing $B_0$ field generated by the OL polepiece is varied, the signal shifts according to the gyromagnetic ratio of the specimen, following the relation: $\omega = 2\pi \Gamma B_0$. Based on our previous results \cite{jarovs2024electron}, we can evaluate the gyromagnetic ratio of the specimen as $\Gamma = (27.9 \pm 1.2)$ GHz/T, which matches well with literature values $\Gamma = \frac{1}{2\pi}\frac{g_e \mu_B}{\hbar}$, where $g_e$ is the electron's g-factor, $\mu_B$ is the Bohr magneton and $\hbar$ is reduced Planck constant. 

\section*{Conclusion}

In this study, we successfully developed the SPINEM technique, integrating a customized ESR setup with a modified TEM sample holder and specially optimized TEM settings. Our method enables the dynamic excitation and detection of spin systems, providing detailed spectroscopic information on microscopic samples, and allowing for mapping of specimen-induced dynamic magnetic fields with a spatial resolution of 30 µm, comparable to other spin mapping techniques, such as Brillouin light scattering \cite{sebastian2015micro, wang2023deeply, wojewoda2023observing}. At resonance, we show a deflection of $\gamma = 7.5$ nrad due to spin-electron interaction, with a detection sensitivity reaching down to 280 prad. Incorporating advanced STEM techniques such as differential phase contrast or 4D-STEM could further enhance the SPINEM signal and push the spatial resolution towards the atomic scale \cite{Haslinger2024}.

Further adjustments to the setup can also increase the obtainable signal: at 4.7 GHz and 290 K, a spin polarization $\propto ~\omega^2/T$ of only $\sim 4\cdot 10^{-4}$ contributes to SPINEM. 
Lowering the sample temperature to the mK range can dramatically increase the signal. If a similar deflection sensitivity can be maintained for a 100 pm electron probe, even single spin contributions could become accessible in electron microscopy studies \cite{Haslinger2024}.

The results presented herein highlight the potential of SPINEM for various research fields. One potential application is the mapping of magnons and collective spin excitations in 2D layered thin films and nanostructures which could enable atomic-scale investigations relevant to spintronics and magnonics \cite{liu2025correlated}. Our technique also holds promise in the analysis of electron-induced radiation damage and the probing of sensitive biological samples without direct electron-specimen interaction. Finally, the position-encoded temporal information could allow time-resolved studies far beyond the capabilities of conventional electron detectors without the use of sophisticated ultra-fast TEM equipment. This work paves the way for the next generation of MW-pump electron-probe methodologies that could revolutionize the study of quantum materials, spin-based information processing, and magnetization dynamics on the nanoscale.

\section*{Acknowledgments}
The authors thank Giovanni Boero, Andrea Pupic, Santiago Beltrán-Romero, Dennis Rätzel, Stefan Nimmrichter, Thomas Spielauer, Matthias Kolb and Thomas Weigner, and the USTEM team for support and fruitful discussions.
PH thanks the Austrian Science Fund (FWF): Y1121, P36041, P35953. This project was supported by the ESQ-Discovery Program 2019 "Quantum Klystron" hosted by the Austrian Academy of Sciences (ÖAW) and the FFG-project AQUTEM.

\section*{Methods}

\textbf{Experiment}: 
The data is collected on a standard FEI Tecnai F20 TEM. To accurately resolve the beam profile in LAD mode, we utilize a 4K CMOS Gatan Rio camera with a high pixel count of 4096 × 4096 and minimum acquisition time of 50 ms. As a specimen, we use a spin-active radical, $\mathrm{\alpha,\gamma}$-bisdiphenylene-$\mathrm{\beta}$-phenylallyl (BDPA) \cite{BOERO_BDPA} of $\sim$110 × 150 × 240 \textmu m$^3$ size. BDPA is widely used in ESR as a benchmark sample due to its stability at room temperature and high spin density, $\rho \approx 1.5 \cdot 10^{27}$ spins/m$^3$.

For the SPINEM measurement, a CW MW-pump and electron-probe scheme is employed, with the electron beam precisely positioned aloof to the edge of the BDPA specimen. The CW frequency is set to $\nu = 4.89$~GHz, while the magnetic field is finely tuned by varying the OL excitation in steps of 0.0001\% (2.2 µT). However, sweeping the OL excitation disrupts the beam (e.g. parallelism). To compensate for this, we adjust the condenser lens excitation.

\textbf{Image Processing}: 
Each data point consists of four individual images, each with a 5-second exposure time, recorded at a TEM camera length of 600 m in LAD mode. To mitigate slow beam drifts, a diffraction shift correction was applied in between subsequent images, utilizing a center of mass (COM) analysis. In total, each spectrum in Fig. \ref{fig:3} represents 976 individual images, amounting to a total data size of 64 GB. 

To process the individual images, we first apply a 3$\times$3 median filter and a global threshold to suppress camera noise. Next, a COM method is used to refine the center position with sub-pixel precision. Assuming that the deflections $\boldsymbol{\alpha}$, $\boldsymbol{\beta}$, and $\boldsymbol{\gamma}$ at MW phases of $0, \pm2\pi, \dots$ have the same magnitude but opposite directions at $\pm\pi, \dots$, this COM acts as a robust reference point. Relative to this center, we perform a principal component analysis (PCA) to evaluate the pattern tilt $\varepsilon$. For more detail, see SI~\hyperref[sec:AppendixC]{c}. Each data point represents the average of four $\varepsilon$ values, which are determined from individual images. 

Using $\varepsilon$ estimated via the PCA technique, we can further extract the electron pattern length change by bisecting the image at its global COM along the angle $\varepsilon+90$ and performing a COM evaluation for each segment. Averaging the COM positions of both segments yields an intensity-weighted pattern length.

\textbf{Signal Fitting}
The observed offset in our analysis results from sweeping the OL excitation to tune the spin transition across the resonance. This adjustment induces a slight rotation and magnification of the LAD image, which in our setup can only be corrected during post-processing. To account for this, a second degree polynomial (offset) combined with a Gaussian (absorption signal) and Gaussian derivative (dispersion signal) fit is performed to extract the deflections of the electron beam caused by spin precession.

\bibliographystyle{naturenano}
\bibliography{main}

\clearpage

\clearpage

\section*{SI A: Microresonator Power Delivery}\label{sec:AppendixA}
\beginsupplement

\begin{figure}[t]
    \centering
    \includegraphics[width=0.48\textwidth]{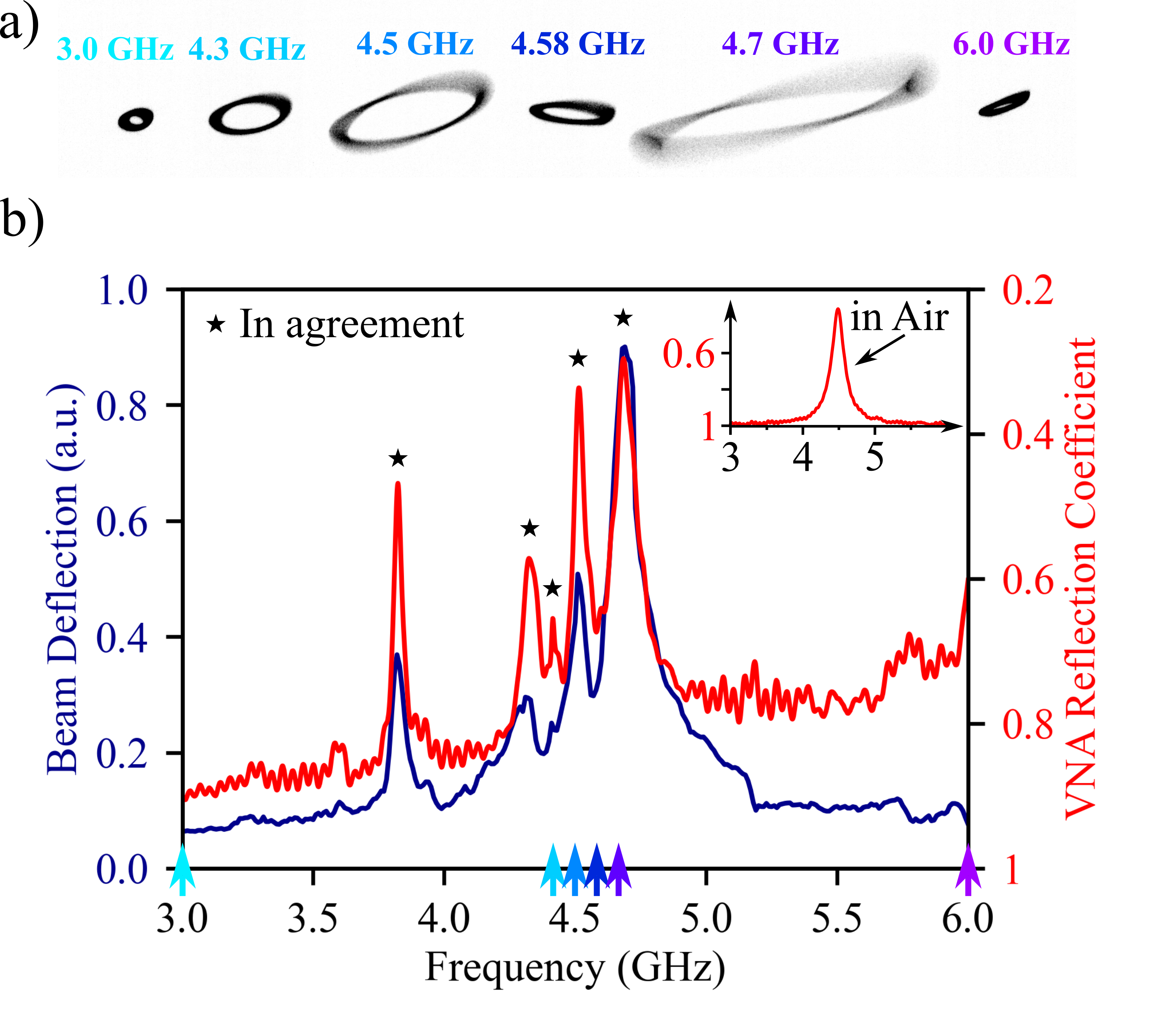}
    \caption{
    \justifying Microresonator impedance match measurement. Different MW driving frequencies deliver varying power to the microresonator, resulting in changes of the $\mathbf{B}_1$ field and beam deflection $\boldsymbol{\alpha}$ as a function of frequency (shown in blue). Maximum beam deflection occurs at the microresonator's impedance match. For comparison, power delivery was also measured conventionally using a vector network analyzer (in red). Both measurements show good agreement, although intensity variations are observed. The inset shows the VNA measurement of the microresonator conducted in air, where additional peaks observed in the TEM environment are attributed to parasitic capacitances and additional resonance modes induced by the TEM environment, acting as a metallic cavity.}
    \label{fig:vna_tem}
\end{figure}

The introduction of the dynamic MW field $\mathbf{B}_1(\omega,t)$ results in a deflection of the electron beam. The beam profile appears as a line or an ellipse (Fig. \ref{fig:vna_tem}a), depending on the linear or elliptical polarization of the $\textbf{B}_1$ field created by the microresonator. At the microresonator impedance match, the $\textbf{B}_1$ field is highly homogeneous within the spatial constraints of the microresonator, resulting in a pattern that appears as a line. The beam deflection $\boldsymbol{\alpha}$ follows:
\begin{equation}
\boldsymbol{\alpha} (\omega, t) \sim \mathbf{v}_{\text{e}} \times \mathbf{B}_{1,\text{max}}(z,\omega) \cos(\omega t)
\end{equation}
where $\mathbf{v}_{\text{e}} = (0,0,-v_{\text{e}})$ is the velocity vector of the electron beam. We neglect transverse velocity components. $\mathbf{B}_{1,\text{max}}(z,\omega) \cos(\omega t)$ is the oscillating magnetic field with a vector magnitude of $\mathbf{B}_{1,\text{max}} = (B_{1,\text{max}},0,0)$. Note that no significant transverse spatial dependence of $B_1$ can be observed, as the microresonator spans a large (>1 mm) section of the specimen region and generates relatively homogeneous fields within its spatial constraints.
We can approximate the total angular deflection of the electron beam by integrating the Lorentz force acting on the electrons over their flight path:
\begin{equation}
\boldsymbol{\alpha} (\omega, t) \approx - \frac{e}{m_{\text{e}}^\star v_{\text{e}}^2} \int_{-\infty}^{+\infty} \mathbf{v}_{\text{e}} \times \mathbf{B}_{1,\text{max}}(z,\omega) \cos(\omega t)\,dz.
\end{equation}
Here, $e$ and $m_{\text{e}}^\star$ are electron charge and relativistic mass and $\boldsymbol{\alpha} (\omega, t) = (0,\alpha,0)$. For the calculations, we assume a homogeneous and linearly polarized $B_1$ field within the spatial extent $l = 1.4$ mm of the microresonator, and zero elsewhere, therefore the $\alpha$ amplitude follows:
\begin{equation}
\alpha (\omega, t) = \frac{e l}{m_{\text{e}}^\star v_{\text{e}}} B_{1,\text{max}}\cos(\omega t).
\end{equation}
The deflection $\alpha$ reflects the power delivered by the MWs from the microresonator, analogous to $B_1$. By analyzing the beam pattern spread, we can evaluate the power delivery and assess the impedance match of the microresonator, see Fig.~\ref{fig:vna_tem}b, blue curve. 

For comparison, the impedance matching was also measured using a vector network analyzer (VNA), see Fig. \ref{fig:vna_tem}b, red curve. A VNA evaluates the power delivered to the device under test (DUT) against the power reflected from it, with reflection coefficient of 0 representing a perfect impedance match, where all power is transferred to the DUT.

The impedance match of our microresonator alone is shown in the inset of Fig. \ref{fig:vna_tem}b. When the microresonator is inserted into the holder and placed inside the TEM, additional peaks appear. This is attributed to parasitic capacitances introduced by the proximity of additional metal components to the microresonator’s conductors. Moreover, the TEM itself acts as a metallic cavity, introducing further resonance conditions. The higher background observed in the VNA measurements inside the TEM is likely due to calibration inconsistencies, as the VNA calibration procedure is performed in air before insertion of the holder into the TEM. This movement of components after the calibration procedure can degrade calibration accuracy.

Overall, both the TEM beam deflection analysis and the VNA measurements show good agreement. Intensity variations arise as the VNA captures the power delivered to the entire DUT, whereas the beam measurement is only locally sensitive to the $B_1$ field, theoretically enabling highly spatially resolved mapping of power delivery to specific regions of the DUT.

\clearpage

\section*{SI B: Beam deflection modeling}\label{sec:AppendixB}

For $\mathbf{B_1}(\omega \approx \omega_{\text{res}} )$, the specimen's magnetization vector $\mathbf{M}(\omega,t)$ begins to precess, resulting in static out-of-plane component $\mathbf{s}(\omega)$ and dynamic in-plane component $\mathbf{m}(\omega,t)$, see Fig. \ref{fig:1}. The electron beam deflection can be modeled by determining this in-plane magnetization of the specimen, achieved by solving the Bloch equation. The in-plane magnetization vectors can be described as:
\begin{equation}
m^{\parallel \text{B}_1} (\omega, t) = M_{x'}(\omega) \cos(\omega t) - M_{y'}(\omega) \sin(\omega t),
\end{equation}
\begin{equation}
m^{\perp \text{B}_1} (\omega, t) = M_{y'}(\omega) \cos(\omega t) + M_{x'}(\omega) \sin(\omega t),
\end{equation}
Here, $m^{\parallel \text{B}_1}$ and $m^{\bot \text{B}_1}$ denote the components of the in-plane magnetization vectors that are, respectively, collinear and perpendicular to the linear polarization of the driving magnetic field $B_1(\omega,t)$. $M_{x'}$ and $M_{y'}$ are the magnetization components in a rotating frame of reference around the $B_0$-axis at the frequency of the driving microwaves, expressed as:
\begin{equation}
M_{x'} (\omega) = \frac{(\omega - \omega_{\text{res}}) (2\pi\Gamma) (\frac{B_1}{2}) T_{2}^2}{1 + [(\omega - \omega_{\text{res}}) T_{2}]^2 + (2\pi \Gamma)^2 (\frac{B_1}{2})^2 T_1 T_2} M_0,
\end{equation}
\begin{equation}
M_{y'} (\omega) = \frac{(2\pi\Gamma)(\frac{B_1}{2}) T_{2}}{1 + [(\omega - \omega_{\text{res}}) T_{2}]^2 + (2\pi \Gamma)^2 (\frac{B_1}{2})^2 T_1 T_2} M_0,
\end{equation}
where $\Gamma = \frac{1}{2\pi}\frac{g_e \mu_B}{\hbar}$ is the gyromagnetic ratio of the sample, and $T_1$ and $T_2$ are the corresponding relaxation times in the parallel and transverse directions relative to the static spin alignment. In conventional ESR, the components $M_{x'}(\omega)$ and $M_{y'}(\omega)$ are measured to determine the parameters $\omega_{\text{res}}$, $\Gamma$, $T_1$, and $T_2$, which provide valuable information about the specimen. The measurements of $M_{x'} (\omega)$ and $M_{y'}(\omega)$ for varying frequency or bias magnetic field $B_0$ across the resonance represent the ESR dispersion and absorption spectra, respectively.

In comparison to the main text, it is apparent that $m(\omega) = \sqrt{M_{x'}^2(\omega) + M_{y'}^2(\omega)}$ and $\theta(\omega) = -\arctan(\frac{M_{x'}(\omega)}{M_{y'}(\omega)})-90^\circ$. Therefore, $m^{\parallel \text{B}_1}(\omega,t) = m \cos(\omega t + \theta)$ and $m^{\bot \text{B}_1}(\omega,t) = m \sin(\omega t + \theta)$.

We consider a spin-active specimen modeled as a point source, with a point-like electron probe positioned nearby. The electron beam is deflected by the dynamic magnetic near-field $\mathbf{B}_{\text{dyn}} (\mathbf{r}, \omega, t)$, generated by the dynamic magnetization $\mathbf{M}(\omega,t)$ of the specimen:
\begin{equation}
\mathbf{B}_{\text{dyn}} (\mathbf{r}, \omega, t) = \frac{\mu_0}{4 \pi r^3} \left[ \frac{3 (\boldsymbol{\mu} \cdot \mathbf{r}) \mathbf{r}}{r^2} - \boldsymbol{\mu} \right],
\end{equation}
where $\boldsymbol{\mu} (\omega,t) = \mathbf{M}(\omega,t)V$ is the dipole moment of the sample, $V$ is the expected sample volume, and $\mathbf{r} = (x, y, z)$ represents the electron probe position coordinates relative to the specimen. The $xy$-plane indicates the sample plane, while the $z$-axis aligns with the TEM optical axis and with the $B_0$ field.

As electrons interact with the sample, the Lorentz force causes an angular deflection of the electron beam:
\begin{equation}\label{eq:lorentz}
\mathbf{F} (\mathbf{r}, \omega, t) = -e \mathbf{v}_{\text{e}} \times \mathbf{B}_{\text{dyn}} (\mathbf{r}, \omega, t).
\end{equation}
Here $\mathbf{B}_{\text{dyn}}(\mathbf{r}, \omega, t) = (B_{\parallel,\text{dyn}}, B_{\perp,\text{dyn}}, B_{\text{z},\text{dyn}})$, where $B_{\parallel,\text{dyn}}$ and $B_{\perp,\text{dyn}}$ are the dynamic field components parallel and perpendicular to the alignment of $\mathbf{B}_{1} (\omega,t) = (B_{1},0,0)$. We neglect the Lorentz force from the electric fields, as well as the transverse velocity components of the electron beam, due to the high longitudinal velocity of the 200 keV electrons: $\mathbf{v}_{\text{e}} = (0,0,-v_{\text{e}})$.

Our SPINEM measurements (Fig. \ref{fig:2}-\ref{fig:4}) provide a measure of angular deflection of the electron beam, obtained by integrating the Lorentz force (eq. \ref{eq:lorentz}) acting on the electrons over their flight path \cite{Haslinger2024}:
\begin{equation}
    \boldsymbol{\delta}_{\text{dyn}} (\mathbf{R}, \omega, t) \approx 
    \frac{-e}{m_{\text{e}}^\star v_{\text{e}}^2} 
    \int_{-\infty}^{+\infty} 
    \mathbf{v}_{\text{e}} \times \mathbf{B}_{\text{dyn}} (\mathbf{r}, \omega, t) \,dz.
\end{equation}
$\boldsymbol{\delta}_{\text{dyn}} (\mathbf{R}, \omega, t) \approx (\gamma, \beta, 0)$ can be interpreted as an average measure of the dynamic magnetic fields, $B_{\parallel,\text{dyn}}$ and $B_{\perp,\text{dyn}}$. The resulting beam deflection due to specimen-electron interaction follows:
\begin{equation}\label{eq:beta}
\beta (\mathbf{R}, \omega, t) = \mathcal{N} \frac{ y^2 m^{\parallel \text{B}_1} - 2 x y m^{\bot \text{B}_1} -  x^2 m^{\parallel \text{B}_1}}{(x^2 + y^2)^2},
\end{equation}
\begin{equation}\label{eq:gamma}
\gamma(\mathbf{R}, \omega, t) = \mathcal{N} \frac{ y^2 m^{\bot \text{B}_1} + 2 x y m^{\parallel \text{B}_1} -  x^2 m^{\bot \text{B}_1}}{(x^2 + y^2)^2}
\end{equation}
where $\beta (\mathbf{R}, \omega, t)$ and $\gamma (\mathbf{R}, \omega, t)$ represent deflection parallel and orthogonal to the deflection $\alpha (\omega,t)$, which is caused by the driving $B_1$ field. $\mathcal{N} = \frac{-e}{m_{\text{e}}^\star v_{\text{e}}}\frac{\mu_0V}{2\pi}$ represents a pre-factor composed of various constants. Taking eq. \ref{eq:gamma} as an example, it can be rearranged as a linear combination of absorption, $M_{y'}(\omega)$, and dispersion, $M_{x'}(\omega)$, spectra:
\begin{equation}\label{eq:gamma_long}
\begin{aligned}
\gamma(\mathbf{R},\omega, t) 
=&\mathcal{N} \Big[ \mathcal{A}(\mathbf{R}) M_{y'}(\omega) + \mathcal{B}(\mathbf{R}) M_{x'}(\omega) \Big] \cos{(\omega t)} \\
&+ \Big[ \mathcal{A}(\mathbf{R}) M_{x'}(\omega) - \mathcal{B}(\mathbf{R}) M_{y'}(\omega) \Big] \sin{(\omega t)},
\end{aligned}
\end{equation}
where $\mathcal{A}(\mathbf{R}) = \frac{y^2-x^2}{(x^2+y^2)^2}$ and $\mathcal{B}(\mathbf{R}) = \frac{2xy}{(x^2+y^2)^2}$. As a consequence, $\gamma$ can therefore be interpreted as a measure of conventional absorption and dispersion spectra. Rearranging eq. \ref{eq:gamma_long} into its simplest form, it can be shown that $\gamma$ represents a sinusoidal deflection of the electron beam:
\begin{equation}\label{eq:gamma_shortest}
\gamma(\mathbf{R},\omega, t) = \mathcal{N} m(\omega) \mathcal{C} (\mathbf{R}) \sin{(\omega t + \theta(\omega) + \phi(\mathbf{R}))},
\end{equation}
where $\mathcal{C} (\mathbf{R}) =\frac{1}{R^2}$ and $\phi(\mathbf{R}) = \arctan{(\frac{2xy}{y^2 - x^2})}$. $\phi(\mathbf{R})$ can be interpreted as an additional phase lag.

Across the resonance, both $\beta$ and $\gamma$ reach a maximum amplitude of $\sim10$ nrad,  proportional to $10^{-4}$ spin polarization at 300 K and $B_0=0.17$T. To resolve $\gamma$ experimentally, we would require nrad sensitivity of the electron beam deflection as well as a time resolution better than $1/4.7~\text{GHz}^{-1} \approx 210~\text{ps}$ at $\nu=4.7$ GHz. These difficulties are mitigated by referencing the measurements to $\alpha$, which acts as a stable phase reference, while creating a position-encoded time resolution.

Therefore, we focus on the change in the electron pattern tilt $\varepsilon$ as the $B_0$ field is swept across the resonance:
\begin{equation}\label{eq:epsilon}
    \tan\varepsilon (\mathbf{R}, \omega,t) = \frac{\gamma (\mathbf{R}, \omega,t)}{\alpha (\omega,t) + \beta(\mathbf{R}, \omega,t) } \approx \frac{\gamma (\mathbf{R}, \omega,t)}{\alpha (\omega,t) }.
\end{equation}
The electron beam pattern, acquired over 5 seconds, is an average of many precession periods ($5~\text{s} \, \cdot \, 4.7 \, \text{GHz}=23.5 \,\cdot \, 10^9$). To the acquired images, we apply the PCA technique for angle estimation. PCA works by averaging over the coordinates of each image and subsequently estimating a direction of maximum variance. For each image, corresponding to a specific $\omega$, this can be reflected in equations by time-averaging beam deflections over a single precession cycle from $0$ to $2\pi$, with phase $\omega t$:
\begin{equation}\label{eq:epsilon_integral}
\tan \varepsilon (\mathbf{R}, \omega) =  \frac{1}{2\pi}\int_{0}^{2\pi} \frac{\gamma (\mathbf{R}, \omega,t)}{\alpha (\omega,t) } \, d(\omega t).
\end{equation}
Using $\alpha(\omega, t) = \alpha_{\text{max}}\cos{(\omega t)}$ and substituting eq. \ref{eq:gamma_long} into eq. \ref{eq:epsilon_integral}, we obtain:
\begin{equation}
\begin{aligned}
\tan \varepsilon& (\mathbf{R}, \omega) 
 = \frac{\mathcal{N}}{2\pi}\int_{0}^{2\pi}\Big\{  
\Big[ \frac{\mathcal{A}(\mathbf{R})}{\alpha_{\text{max}}} M_{y'}(\omega)  
+ \frac{\mathcal{B}(\mathbf{R})}{\alpha_{\text{max}}} M_{x'}(\omega) \Big] \\
&\quad+ \Big[ \frac{\mathcal{A}(\mathbf{R})}{\alpha_{\text{max}}} M_{x'}(\omega)
- \frac{\mathcal{B}(\mathbf{R})}{\alpha_{\text{max}}} M_{y'}(\omega) \Big]  
\tan{(\omega t)}  
\Big\} \, d(\omega t).
\end{aligned}
\end{equation}
Since $\int_{0}^{2\pi} \tan{\omega t} \,d(\omega t) = 0$, we get:
\begin{equation}\label{eq:angle_final}
\tan \varepsilon (\mathbf{R}, \omega) =\mathcal{N} \Big[ \frac{\mathcal{A}(\mathbf{R})}{\alpha_{\text{max}}} M_{y'}(\omega) 
+ \frac{\mathcal{B}(\mathbf{R})}{\alpha_{\text{max}}} M_{x'}(\omega) \Big]
\end{equation}
In this way, the time dependence of $\varepsilon$ is eliminated and we obtain a static measure of specimen properties. 

In Fig. \ref{fig:3}-\ref{fig:4}, we use eq. \ref{eq:angle_final} to compute the electron beam deflection $\gamma_{\text{max}}$ caused by the interaction with the spin system: $\gamma_{\text{max}} = \alpha_{\text{max}}\tan{\varepsilon}$. Consequently, our measurements directly correspond to the $\gamma$ measurement at a MW phase $\omega t = 0,\pm2\pi, \dots$:
\begin{equation}
\gamma_{\text{max}}(\mathbf{R},\omega, t)
= \mathcal{N} \Big[\mathcal{A}(\mathbf{R}) M_{y'}(\omega) + \mathcal{B}(\mathbf{R}) M_{x'}(\omega) \Big].
\end{equation}
This measurement is influenced only by the position of the electron probe $\mathbf{R}$ and the $B_0$ field, which determines whether the resonance condition is satisfied. To demonstrate the position dependency, calculated 2D deflection maps of $\gamma$ are shown in Fig. \ref{fig:3}d-f, at -140 mT, 0 mT and +140 mT detuning from resonance, respectively.

For the special case where $x = 0$ and $y \neq 0$, which reflects the electron beam position (1) in Fig.~\ref{fig:3}, the calculated deflection is given by:
\begin{equation}
\gamma_{\text{max}} (\mathbf{r, \omega}) = \frac{\mathcal{N}}{y^2} M_{y'}(\omega),
\end{equation}
\begin{equation}
\beta_{\text{max}}(\mathbf{r}, \omega) = \frac{\mathcal{N}}{y^2} M_{x'}(\omega).
\end{equation}
Similarly, for $x \neq 0$ and $y = 0$, reflecting position (3) in Fig.~\ref{fig:3}, 
\begin{equation}
    \gamma_{\text{max}} (\mathbf{r, \omega}) =- \frac{\mathcal{N}}{x^2} M_{y'}(\omega), 
\end{equation}
\begin{equation}
\beta_{\text{max}}(\mathbf{r}, \omega) = -\frac{\mathcal{N}}{x^2} M_{x'}(\omega).
\end{equation}
A change in beam position from (1) to (3) is equivalent to a MW phase shift of $\pi$ at the first mixing stage as utilized in conventional ESR. For $x = y$, reflecting beam position (2) in Fig. \ref{fig:3}, a measurement of $\gamma$ results in a measurement of an ESR dispersion spectrum:
\begin{equation}
\gamma_{\text{max}} (\mathbf{r}, \omega) = \frac{\mathcal{N}}{2x^2} M_{x'}(\omega), 
\end{equation}
\begin{equation}
\beta_{\text{max}} (\mathbf{r},\omega) = -\frac{\mathcal{N}}{2x^2} M_{y'}(\omega).    
\end{equation}

These equations demonstrate a complex, yet predictable interplay 
of beam deflections $\beta$ and $\gamma$ and the electron probe position $\mathbf{R}$. At other positions, the measured deflections represent a linear superposition of absorption and dispersion spectra, see eq. \ref{eq:beta},\ref{eq:gamma}.

\clearpage

\section*{SI C: Principal Component Analysis}\label{sec:AppendixC}

To determine the orientation of the electron beam pattern using PCA, we first compute the global COM of the image, weighted by pixel intensities $ I_i $:
\begin{equation} 
x_{\text{COM}} = \frac{\sum I_i x_i}{\sum I_i}, \quad 
y_{\text{COM}} = \frac{\sum I_i y_i}{\sum I_i}.
\end{equation}
The image coordinates are then shifted relative to this weighted COM:
\begin{equation} 
X = x_i - x_{\text{COM}}, \quad Y = y_i - y_{\text{COM}}.
\end{equation}
Next, we compute the intensity-weighted covariance matrix:
\begin{equation} 
\mathbf{C} =
\frac{1}{\sum I_i} 
\begin{bmatrix}
\sum I_i X^2 & \sum I_i X Y \\
\sum I_i X Y & \sum I_i Y^2
\end{bmatrix}.
\end{equation}
We then compute the eigenvalues ($\lambda$) and eigenvectors ($v$) by solving:
\begin{equation}
\mathbf{C} \boldsymbol{v} = \lambda \boldsymbol{v}.
\end{equation}
The two eigenvectors $ \boldsymbol{v}_1 $ and  $\boldsymbol{v}_2 $ correspond to the principal directions of the distribution, where each eigenvector has components:
\begin{equation}
\boldsymbol{v}_i = \begin{bmatrix} v_{i,1} \\ v_{i,2} \end{bmatrix}.
\end{equation}
The eigenvector corresponding to the largest eigenvalue, $ \boldsymbol{v}_{\max} $, defines the major axis of the beam pattern. The orientation angle of the beam pattern is then given by:
\begin{equation}
\tan\varepsilon= (\frac{v_{\max,2}}{v_{\max,1}}),
\end{equation}
where  ($v_{\max,1}, v_{\max,2}$) are the x and y components of the principal eigenvector.

\end{document}